\documentclass{pnastwo}

\usepackage{graphicx}

\usepackage{amsmath}
\usepackage{amssymb}

\renewcommand\thetable{S\arabic{table}}
\newcommand{\wh}{\hat}

\contributor{Submitted to Proceedings
of the National Academy of Sciences of the United States of America}
\url{www.pnas.org/cgi/doi/10.1073/pnas.XXXXXXXXXX}
\copyrightyear{2014}
\issuedate{Issue Date}
\volume{Volume}
\issuenumber{Issue Number}

\begin{document}


\title{Calorimetric glass transition in a mean field theory approach}




\author{Manuel Sebastian Mariani\affil{1}{D\'epartement de Physique,
Universit\'e de Fribourg, Ch. du Mus\'ee 3, CH-1700 Fribourg, Suisse}, 
Giorgio Parisi\affil{2}{Dipartimento di Fisica,
Sapienza Universit\`a di Roma, P.le A. Moro 2, I-00185 Roma, Italy}\affil{3}{INFN, Sezione di Roma I, IPCF -- CNR, P.le A. Moro 2, I-00185 Roma, Italy},
\and Corrado Rainone\affil{2}{Dipartimento di Fisica,
Sapienza Universit\`a di Roma, P.le A. Moro 2, I-00185 Roma, Italy}\affil{4}{LPT,
\'Ecole Normale Sup\'erieure, UMR 8549 CNRS, 24 Rue Lhomond, 75005 Paris, France.}}

\contributor{Submitted to Proceedings of the National Academy of Sciences
of the United States of America}

\maketitle

\begin{article}

\begin{abstract} 
The study of the properties of glass-forming liquids
is  difficult for many reasons. Analytic solutions of mean field models are usually available only for systems
embedded in a space with 
an unphysically high number of spatial dimensions; on the experimental and numerical side, the study of the properties of 
metastable glassy states requires to thermalize the system in the 
supercooled liquid phase, where the thermalization time may be extremely large. 
We consider here an hard-sphere mean field model which is solvable in any number of spatial 
dimensions; moreover we easily obtain thermalized configurations even in the glass phase. 
We study the three dimensional version of this model and we  perform Monte Carlo simulations which mimic heating and cooling experiments performed on ultra-stable glasses. The 
numerical findings are in good agreement with the analytical results and qualitatively capture the features of ultra-stable 
glasses observed in experiments. 
\end{abstract}

\keywords{glass transition|mean-field theory|ultra-stable glasses|planting|replica theory|complexity}





\noindent \noindent\rule{8cm}{0.4pt}
\\
{\bf Significance} 

Understanding the properties of glasses is one of the major open challenges of theoretical physics. 
Making analytical predictions is usually very difficult
for the known glassy models. Moreover, in experiments and numerical simulations thermalisation of glasses 
cannot be achieved without sophisticated procedures, like the vapour deposition technique.
In this work we study a glassy model which is simple enough to be analytically solved and which can be thermalised in the glassy phase with a simple
numerical method, opening the door to the intensive comparison between replica theory predictions and numerical outcomes.




\noindent\rule{8cm}{0.4pt}

\section{Introduction}

The theoretical interpretation of the properties of glasses is highly debated. There are two extreme viewpoints:
\begin{itemize}
\item One approach, the Random First Order Transition (RFOT) theory \cite{KTW}, 
which uses mostly the replica method \cite{MPV} as its central tool, assumes that the dynamical properties of glasses do reflect the properties
of the appropriate static quantities (like the Franz-Parisi potential \cite{FP}): for a review see \cite{MPV,zamponi}.
\item The other approach (Kinetically Constrained Models, (KCMs) assumes 
that the glass transition is a purely dynamical phenomenon without any counterpart in static quantities \cite{scl4p,RS,CG}.
\end{itemize}
The mean field version the RFOT approach predicts the presence of a \emph{dynamical transition} (identified with the Mode-Coupling 
transition \cite{MCT}) at a nonzero temperature $T_d$, whereupon the configuration space of the glass-former
splits into a collection of metastable states. 
Below $T_d$, the system will remain trapped inside a metastable state. Beyond mean field theory the dynamical transition $T_d$ becomes a cross over point: at $T_d$ the correlation time and the dynamical correlation length become very large, but finite. Below the $T_d$ the dynamical correlation time becomes very large and 
 it becomes comparable to the human timescales, leading to the phenomenological glass transition. 
In the  KCM approach  the glass transition is a phenomenon originated only by constraints on the dynamics,  while
the RFOT picture views the off equilibrium states as metastable, \emph{thermodynamic states}, they can be identified with the minima of a suitable {\em equilibrium} free-energy functional
and can then be studied using a modified equilibrium formalism, generally built on the replica method.

According to replica formalism, the system explores the whole collection of possible states, 
with lower and lower free-energy, as the temperature is lowered from $T_d$ to another temperature $T_K$ 
(the Kauzmann temperature) where the states with the lowest free energy are reached.
Most RFOT models (but actually not all, since $T_K=0$ for some models) predict then an equilibrium phase transition at $T_K$,
with a real divergence of the relaxation time.

To test this scenario, it would be necessary to perform experiments and simulations at various temperatures
in this range, but then one must face the problem of equilibrating the glass-former at temperatures
$T\approx T_K \ll T_g$ (where $T_g$ is the phenomenological glass transition temperature), where it is by definition impossible to do so. Indeed, a simple estimate shows 
that the increase of the equilibration time below $T_d$ is so sharp that one cannot get nearer to $T_K$ than $\Delta T \approx \frac{1}{3}T_K$ 
without falling out of equilibrium, making for us impossible to get a good look at the lowest states: only the high free-energy
states near $T_d$ can be probed experimentally.

Some progress in this direction has been made recently both in experiments \cite{experimental} and numerical simulations 
\cite{ultrastable}, with the introduction of the so-called vapor deposition technique, which allows one to obtain extraordinarily stable 
glasses (usually referred to as ultrastable glasses \cite{ultrastable, sciortino,ultrastable1,ultrastable2}) in a relatively short time, even for temperatures 
much lower than $T_d$. First numerical simulations on an ultrastable glass of binary Lennard-Jones mixture seem
to support the existence of a thermodynamic phase transition \cite{ultrastable}. On the theoretical side, the intrinsic out-of-equilibrium nature of glass poses another challenge, 
since the methods of equilibrium statistical mechanics cannot be used in the usual way, requiring, in principle, to resort to dynamical tools. 
This strategy is actually viable, and was used for example by Keys et. al in \cite{KGC}, where a suitably 
tuned East-model has been shown to reproduce well the experimental behavior observed in DSC (Differential Scanning Calorimetry) 
experiments on different glass-former materials, for example the Glycerol \cite{glycerol} and the Boron Oxide \cite{boronoxide}. 
This approach however has the drawback of being phenomenological in nature.

The recent introduction \cite{mk} of a semi-realistic soluble model for
glasses (the Mari-Kurchan model, MK) gives us the possibility to address both the equilibration and the theoretical problem. It allows us to obtain equilibrated configurations also beyond the dynamical transition
 and deep into
the glass phase, using the so-called \emph{planting} method \cite{csp}. 
Moreover, it is in principle solvable in the replica method, allowing us to study the
metastable glassy states with a \emph{static} formalism, without having to solve the dynamics.

Our aim is to use this model to simulate slow annealing experiments usually performed on
glasses and ultrastable glasses,
in order to compare the numerical outcomes with experimental results and theoretical predictions in the replica method.

\section{The model}
\label{model}

We consider the potential energy of the family of models introduced by Mari and Kurchan (MK model) \cite{mk}:
\begin{equation}
  V(\underline{x},\underline{l})=\sum_{(i,j)}v(\mathbf{x_{i}-x_{j}-l_{ij}}),
 \label{marikurchan}
\end{equation}
where $\underline{x}=\{\mathbf{x_{1}},\dots,\mathbf{x_{N}}\}$ are $N$ $d$-dimensional vectors, representing particles positions, and the particles move in a $d$ dimensional cube or size $L$, with periodic boundary conditions. The main feature of the model are the variables
 $\underline{l}=\{\mathbf{l_{ij}}\}$: they  are $N(N-1)/2$ 
quenched random vectors, called \emph{random shifts}, independently drawn out from an uniform probability distribution inside the cube. The function
$v$ could be in principle any interesting short-ranged repulsive pairwise interaction.

The main effects of the random shifts is to destroy the direct 
correlation among the particles that interact with a given particle \cite{mk}. This makes 
the computation of static quantities very simple, because in the Mayer
expansion of the grand-canonical potential only the tree diagrams survive in the thermodynamic 
limit \cite{mk}. The idea is quite old \cite{kraichnan}, it had important application to turbulence, 
but it has only recently been applied to glasses.

\subsection{Static thermodynamic properties in liquid phase}
 
Here we will summarise analytical and numerical results obtained by Mari and Kurchan for this model.
In the following $D$ will denote the diameter of spheres. In hard-sphere systems the potential $v(\mathbf{x})$ is infinite at distances less the $D$ and  the role of inverse temperature
is played by the \emph{packing fraction}
$\varphi=N\mathcal{V}_{d}(D)/L^{d}=\rho\mathcal{V}_{d}(D)$, where $\mathcal{V}_{d}(D)$ is the volume
of the $d$-dimensional sphere of diameter $D$; we will call it density absorbing the multiplicative factor in its definition.
 
The Hamiltonian contains random terms and the interesting quantities have then to be averaged over these parameters.
We can define the  {\it annealed} entropy $S^A$  and the  {\it quenched} entropy $S^Q$ given by
\begin{equation}
 S^A\equiv \log(\overline{Z(\underline{l})})\, , \ \ \    S^Q\equiv \overline{\log(Z(\underline{l})) }\, .\label{entropymft}
\end{equation}
The computation of $S^A$ can be easily done and one finds
\begin{equation}
  s^A(\rho)=\frac{S^A(\rho)}{N}=-\log{(\rho)}-2^{d-1}\,\mathcal{V}_{d}(D)\,\rho+\log{(N)}.
 \label{entropyannealed}
\end{equation}
The presence of the $\log{(N)}$ term is due to the fact that in this model particles are distinguishable for a given realisation of random shifts.

A more interesting quantity is the {\it quenched} entropy. In this model one finds that $S^A(\rho)=S^Q(\rho)$ in the liquid phase,
i.e. below the Kauzmann transition density $\varphi_{k}$. The Kauzmann transition is avoided in the thermodynamic limit: 
the total entropy $s^A$ 
grows as $\log{(N)}$ while the vibrational entropy is a non-decreasing function of $\rho$ that diverges in the infinite-density limit.
This implies that the configurational entropy contains a term proportional to $\log(N)$ and thus the value $\varphi_{K}$ where the 
configurational entropy vanishes diverges logarithmically in the thermodynamic limit.

Using standard termodynamic relations one can derive from \eqref{entropyannealed} the liquid-phase equilibrium equation of state
\begin{equation}
 P=\rho+2^{d-1}\, \mathcal{V}_{d}(D)\,\rho^{2},
 \label{eqstateeq}
\end{equation}
{{where $P$ is the pressure.}}

For what concerns the radial distribution function, one has to take the random shifts into account:
\begin{equation}
g(r)=\frac{1}{\rho^{2}}\,\overline{\left<\sum_{i\neq j}^{N}\delta \left(\left|\mathbf{x_{i}-x_{j}+l_{ij}}\right|-r\right)\right>},
\label{rdf}
\end{equation}
where the bracket average is computed using the ensemble distribution function (Gibbs-Boltzmann distribution at equilibrium)
while the bar average is computed using the random shifts probability distribution.
The result is
\begin{equation}
 g(r)= \theta(r-D)\, ,
  \label{rdfmk}
\end{equation}
where $\theta$ is the usual Heaviside step function. This result is the same obtained with high dimensional hard-spheres \cite{denseamorphousHS}, but the mean-field nature of the model has allowed us to get it in any number of spatial dimensions. The equilibrium pressure is related to density by the usual relation for hard spheres \cite{mcdonald}
\begin{equation}
  P=\rho+2^{d-1}\,\mathcal{V}_{d}(D)\, g(D)\,\rho^{2},
 \label{stateeq}
\end{equation}
from which, using \eqref{rdfmk}, the equilibrium equation of state \eqref{eqstateeq} can be derived again.

\subsection{Glassy properties}

The model is interesting because in spite of the extreme simplicity of the statics (a feature that it has in common with facilitated models) the dynamics is extremely complex. At high densities there is glass phase that in the thermodynamic limit is separated from the liquid phase by a Mode Coupling transition. This transition exists only if we embed the model in a space with an infinite number of dimensions $d$; when $d<\infty$, hopping effects destroy the transition which becomes only a crossover region \cite{parisijin}.\\
Accurate simulations \cite{parisijin} give an higher value for the mode-coupling dynamical density, i.e.  $\varphi_{d}=1.91$. A more careful analysis of the properties of the system near the putative mode-coupling transition can be found in \cite{parisijin}, where the effects of hopping are carefully studied.
Other features, like a violation of Stokes-Einstein relation and dynamical heterogeneities, are present in this model \cite{mk,parisijin}.

\section{Numerical simulations}
\label{numres}

{{When a glass is gradually
heated during DSC experiments}} thermodynamic quantities, like the internal energy, continue to follow the glassy behavior also in the liquid phase, 
until the so-called onset temperature $T_{on}$ is reached. For $T>T_{on}$ the system {{gradually approaches equilibrium; during this relaxation process
the specific heat reaches a maximum value, higher than the equilibrium one.}}
The value of $T_{on}$ quantifies the stability of the initial glass and is considerably higher for glasses prepared through the vapour deposition technique
than ordinary glasses aged for many months \cite{experimental, ultrastable}.
The vapour deposition procedure has been recently mimicked by a computer algorithm, and numerical simulations over a
Lennard-Jones binary mixture showed the same behavior \cite{ultrastable, sciortino}.

{{We aim to study numerically this deviation from equilibrium in the liquid region and the subsequent relaxation process in the MK model.
In the MK model we are able to obtain equilibrium configurations beyond the dynamic transition via a special procedure,
allowed only by the presence of random shifts, 
the so-called \emph{planting} \cite{csp} method. Basically planting consists in two steps: the generation of a random configuration of
sphere positions, independently drawn out from the uniform distribution over the volume, and the generation of the random shifts configuration $\{\mathbf{l}_{ij}\}$
so that the non-overlap condition imposed by the hard-sphere potential energy \eqref{marikurchan} is satisfied
for every pair of spheres (see SI for all details).
The mean-field nature of the 
interaction guarantees that planted configurations are equilibrated \cite{csp}. The planted glass in the MK model, like vapour-deposited ultrastable
glasses in real world, is the best possible starting point for the study of the deviation from equilibrium in the liquid phase.
We start from a planted configuration and mimic respectively heating in DSC experiments on our hard-sphere system by 
running adiabatic step-wise decompression  scans, where the system performs jumps between different density values 
and a large number of Monte Carlo steps for each density value, in order to reach thermalisation. }} 

We refer to SI for all other simulation details.
We present now the results of numerical simulations, based on Monte Carlo method, of a system composed by $N=800$ spheres of diameter $D=1$
in $d=3$ dimensions,
with periodic boundary conditions.

\subsection{Decompression jump and spheres contact region emptying}

\begin{figure}[t]
\includegraphics[width = 0.7\columnwidth,angle=270]{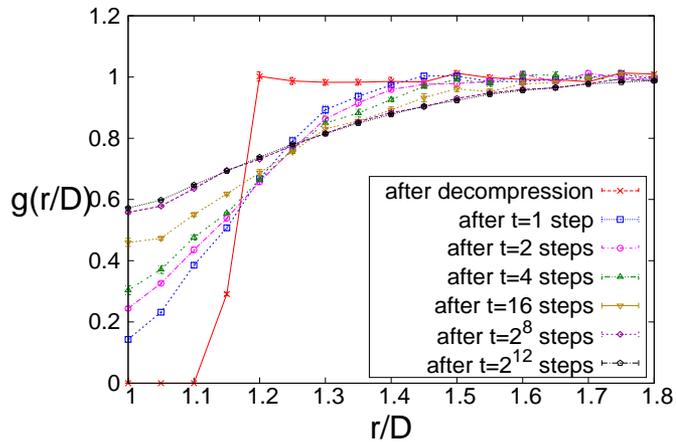}
\caption{Radial distribution function $g(r)$ for $\varphi_{1}=1.7$ at various times $t$,
starting from a thermalised initial configuration at $\varphi_{0}=2.5$. Immediately after the density jump, there are no particles in contact and $g(r)$ shows
a drop where $r\sim D$ (red points). While the system evolves at the new density value $\varphi_{1}=1.7$, spheres gradually return in contact
partially filling the dip. \label{rdf_times}}
\end{figure}

The outcome of the planting technique is a thermalized initial configuration at a certain density $\varphi_{0}>\varphi_{d}$.
We discuss now the effects a density jump $\varphi_{0}\to\varphi_{1}$  on $g(r)$, where $\varphi_{1}<\varphi_{0}$ (decompression).
Results for $\varphi_{0}=2.5$ and $\varphi_{1}=1.7$ are shown in Fig. \ref{rdf_times}.
When spheres radius is decreased, particles originally in contact separate, 
causing a drop of $g(r)$ in the contact region $r\sim 1$.
While the system evolves at the new density value $\varphi_{1}$, gradually particles return in contact, causing
the filling of the contact zone. 
In the glass phase this filling can only be partial for realizable time scales. 
In the liquid region and for densities sufficiently far
from the dynamical transition, it is possible to see a complete filling.

\begin{figure}[t]
\includegraphics[width=0.7\columnwidth,angle=270]{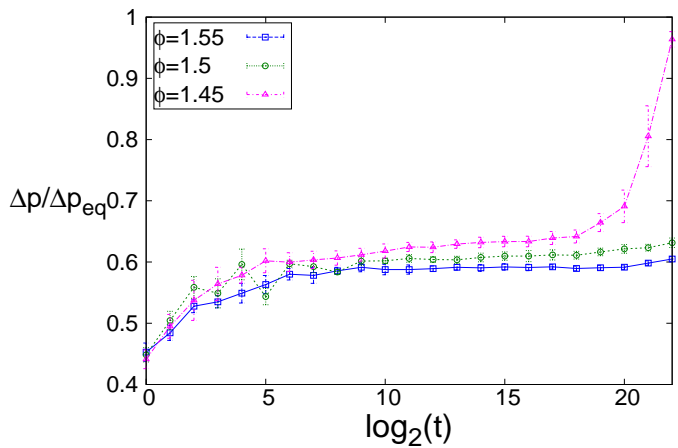}
\caption{Temporal behavior of the ratio $\Delta p/\Delta p_{eq}=g(1)$ between the excess pressure $\Delta p=p-1$ and the equilibrium value $\Delta p_{eq}$
for various $\varphi$ values in the liquid phase, for 
the decompression protocol with $\varphi_{0}=2.5$, $\varphi_{1}=1.6$, $\varphi_{2}=1.55$,
$\varphi_{3}=1.5$, $\varphi_{4}=1.45$. 
{{We do not show the curves for $\varphi=\varphi_{0}=2.5$ (planted configuration, $g(1)=1$) and $\varphi=\varphi_{1}=1.6$, not comparable with
$\varphi=1.55,1.5,1.45$ because the system reaches $\varphi=1.6$ after a larger density jump.}}
Only for $\varphi=1.45$ we can observe a clear sign of structural relaxation, while for higher
densities the system remains trapped in the original metastable state.\label{g1_vs_k}}
\end{figure}

We consider now the following decompression protocol:
we start from a planted configuration at $\varphi_{0}=2.5$, we jump to $\varphi_{1}=1.6$ and wait $2^{22}$ steps, then we jump to
$\varphi_{2}=1.55$ and we wait again $2^{22}$ steps,
then $\varphi_{3}=1.5$ and $\varphi_{4}=1.45$. In Fig. \ref{g1_vs_k} it is shown the temporal behaviour of $g(1)$ for various values of $\varphi$.
We do not see structural relaxation for $\varphi=1.55,1.5$ (liquid phase): the system, 
after a partial, fast relaxation process reaches the metastable plateau and it has not enough time to escape. 
When $\varphi=1.45$, the lifetime of the original metastable state is smaller 
than $2^{22}$ steps and we observe a clear structural relaxation, corresponding to the complete
filling of the contact zone.

\subsection{Mean Square Displacement and structural relaxation time}

To study the behaviour of the relaxation time as a function of density in the liquid phase and evaluate the
dynamic glass transition density $\varphi_{d}$,
we turn our attention to another observable, the mean square displacement $\Delta(t)$ (MSD) of spheres from their initial positions:
\begin{equation}
  \Delta(t)=\frac{1}{N}\sum_{i=1}^{N}|\mathbf{x_{i}}(t)-\mathbf{x_{i}}(0)|^{2}.
\label{eq:msd}
\end{equation}
Since we are interested only in relaxation time, we start from a thermalised configuration at $\varphi_{0}=2.5$ and we jump directly to the density value $\varphi$
we are studying. We stress that in equation \eqref{eq:msd} $\mathbf{x_{i}}(0)$ is physical position of sphere $i$ immediately after the density jump.
We let the system evolve for $2^{k}$ steps at this density $\varphi$. 
For $\varphi<\varphi_{d}$, as expected in a glassy system 
not too far from its dynamical glass transition,
we can observe a two-steps relaxation process. 
The system reaches a metastable plateau after about $2^{10}-2^{11}$ steps, then remains trapped in it for long time,
after which it escapes. 
For each value of density we fitted the MSD's escape from the plateau with a power law function and obtained a value of the relaxation time
(see SI for all details).
We fitted the resulting curve of $\tau_{R}$ as a function of $\varphi$, displayed in Fig. \ref{relaxation}, with the power law behaviour
$\tau_{R}=A(\varphi_{0}-\varphi)^{-\gamma}$,
obtaining in this way $\varphi_{0}=1.73\pm0.04$, $\gamma=4.1\pm0.7$.  
One can notice that the value for $\gamma$ is not too different 
from the one obtained in \cite{mk} performing a similar analysis on the relaxation time, 
while the value of $\varphi_{0}$ is definitely smaller than $\varphi_{d}$, but not too far from the one obtained in \cite{mk}.

\subsection{Decompression and compression scans. Qualitative comparison with  experiments}

\begin{figure}[t]
\includegraphics[width=0.7 \columnwidth,angle=270]{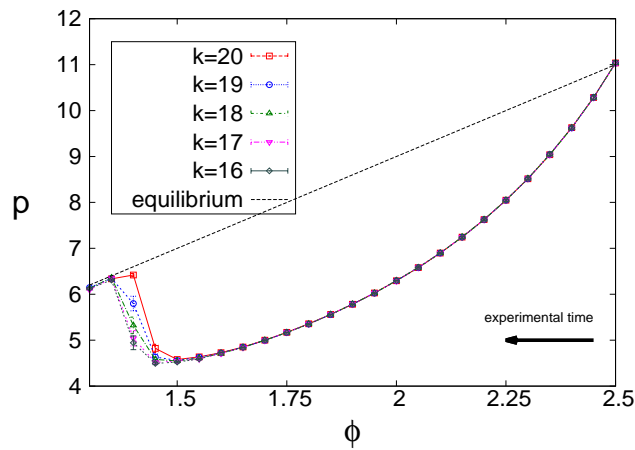}
\caption{Reduced pressure $p$ as a function of density in a decompression protocol with $\varphi_{0}=2.5$, $\Delta\varphi=-0.05$.
Different colors correspond to different decompression rates ($k$ is the number of Monte Carlo steps performed at each density value). \label{g1phi_variousk}}
\end{figure}

In Fig. \ref{g1phi_variousk} we represent the behaviour of the {\emph{reduced pressure $p=P/\rho$}} as a function of density for a
decompression protocol with
starting density $\varphi_{0}=2.5$ and a constant density-jump amplitude $\Delta\varphi=\varphi_{n}-\varphi_{n-1}=\varphi_{1}-\varphi_{0}$. 
We have different curves for different values of the number $2^{k}$ of Montecarlo steps performed at each density value.
We see a deviation from equilibrium independent from the decompression rate for sufficiently high $\varphi$.
For values well above $\varphi_{d}$, for example $\varphi=2.0$,
the system is in the glass phase, so relaxation takes place inside the original metastable state and pressure deviates from its equilibrium value.
This deviation continues for the largest density values below $\varphi_{d}$, for example $\varphi=1.6,1.65,1.7$:
the system continues to relax inside the original metastable state, having yet no sufficient time to reach equilibrium.
When density is sufficiently low, the lifetime of the original metastable state becomes smaller than $2^{k}$ and the onset of relaxation towards equilibrium takes place.
The relatively sharp pressure reclimb is dependent on decompression rate, and it is faster for slower rates.

{{The decompression protocol adopted for our system, composed by hard spheres, is equivalent to the typical DSC's heating scans, 
with two crucial  differences: (a) In DSC experiments we move toward the glassy phase by decreasing the temperature: 
in the case of  hard spheres the inverse of the density plays the same role of the temperature. (b) The starting configurations of the dynamics
are fully equilibrated and this corresponds only to the case of DSC with infinitely slow cooling speed and relatively fast heating speed.}}
The observed deviation of pressure from equilibrium for $\varphi<\varphi_{d}$ is qualitatively the same phenomenon typically observed in {{DSC experiments.}}
One point is important to notice: in DSC experimental {{heating}} data the relaxation toward equilibrium for $T>T_{on}$ 
is gradual and smooth \cite{experimental},  {{internal energy and enthalpy are continous at the onset
point and the specific heat gradually reaches its maximum value}}, while in Fig. \ref{g1phi_variousk}
the reclimbing of pressure seems relatively sharp, {{probably signaling an underlying singularity (with infinite compressibility)}}. 

\begin{figure}[t]
\centering
\includegraphics[width=0.7\columnwidth, angle=270]{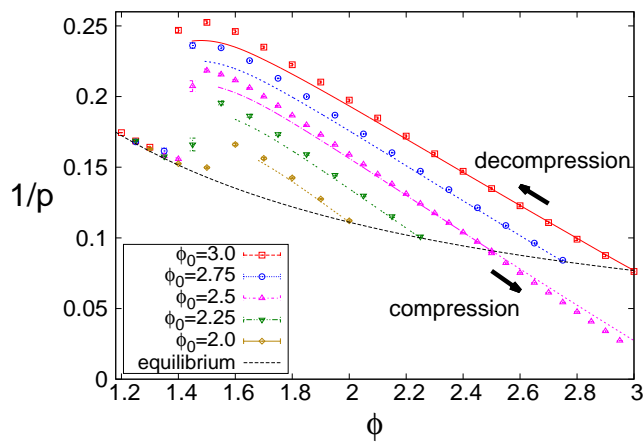}
\caption{Inverse of the reduced pressure $p$ as a function of density during compression and decompression scans.
Different colours correspond to different values of the planting density $\varphi_{0}$.
Points are the results of the Monte Carlo numerical simulations, lines are the analytical results obtained from the replica method within the isocomplexity
assumption, as
described in the text. {{Black line represents the equilibrium pressure (Eq. \eqref{eqstateeq}) 
and black arrows indicate the direction of experimental time, which runs from
right to left during decompression scans (points above the equilibrium line) and from left to right during compression (points below the equilibrium line).}}
\label{comparison}}
\end{figure}

{{
In Fig. \ref{comparison} curves for different values of $\varphi_{0}$ are represented, corresponding to different metastable states.
The performed scans start from a planted configuration, corresponding to a point on the equilibrium line,  
and lead the system to a pressure lower than the equilibrium value during decompression.
As expected, during decompression 
the relaxation of pressure towards equilibrium is sharp and it starts at a lower onset density the higher is $\varphi_{0}$, 
i.e. the more stable is the original glassy configuration: the system has memory of the inital state of the glass (\emph{hysteresis}).
This effect is analogous to what is observed in ultrastable real glasses \cite{experimental}:
the more stable is the initial glass obtained via vapour deposition, the longer is the deviation from equilibrium in the liquid phase
and, as a result, the higher is the onset temperature.
When we compress the system (only from $\varphi_{0}=2.5$ in Fig. \ref{comparison}), we see that the pressure 
becomes higher than the equilibrium one, as expected. This effect mirrors what happens in decompression and the two sets of data
concerning decompression and compression scans from $\varphi_{0}=2.5$ join smoothly, as expected.}}

\section{Replica computation of metastable states curves}
\label{REPLICA}

So far we have shown how the MK model allows to prepare the system in a glass state, 
even at densities much higher that the dynamical one, without incurring in the problem of extremely large equilibration times. 
In addition, this model has another remarkable advantage: it is in principle solvable, thanks to its mean-field nature: the interaction network is tree-like (or alternatively without loops) in the thermodynamic limit, like in Van der Waals liquids \cite{mcdonald,mk},
and thus it also allows for a ready comparison between numerics and analytic computations. 
In particular, it allows us to perform computations in the replica method. Although the MK model is soluble, it actual analytic solution is exceedingly complex \cite{MezardZamponi}, so we have to resort to make some approximation: here we assume that the cages have a Gaussian shape \cite{parisijin}.

In the replica approach to the glass transition \cite{KTW,zamponi}, it is assumed that for densities $\varphi>\varphi_{d}$ the configuration space can be unambigously splitted in subsets, denoted as metastable glassy states.
These states are theoretically identified with the local minima of a suitable functional, which plays in this context 
the same role of the Thouless-Anderson-Palmer (TAP) free energy in spin glasses \cite{MPV}. 
In a mean field situation and in the thermodynamic limit, where metastable states live forever, the system 
becomes then immediately trapped in one of these states and fails permanently to attain relaxation (the so-called Mode Coupling Transition).
On the other hand, out of mean field or with finite system size, the system will be finally able to hop out of the state \cite{parisijin} 
and relax, although an extremely large time will be needed to do so \cite{scl4p}.

The most important feature of these metastable states is that they are degenerate, this is, they can have the same free entropy. In fact, if one fixes a density $\varphi>\varphi_d$ and a value $s$ for the free entropy, it is possible to see that the number of states that share it (in the functional picture, the number of minima which all have the same height $s$) scales exponentially with the size of the system, $\mathcal{N}(s,\varphi) \propto e^{\Sigma(s,\varphi)N}$. This causes the total free entropy of the system to gain an extra term to take into account this fact:
\[
 S(s,\varphi)=s+\Sigma(s,\varphi),
\]
where $\Sigma(s,\varphi)$ is called \emph{complexity} (or alternatively configurational entropy), a central quantity in replica theory.

 \begin{figure}[t]
\centering
\includegraphics[width=0.7\columnwidth, angle=270]{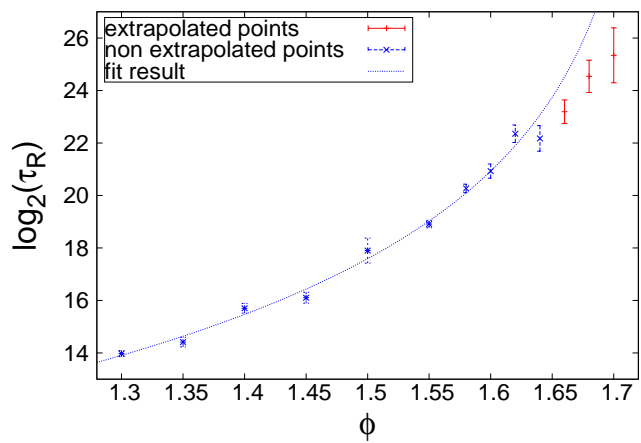}
\caption{Relaxation time as a function of density, computed as described in the text. Red points are the result of an extrapolation 
($\tau_{R}$ larger than the largest number of performed Monte Carlo steps at each density value, i.e. $2^{22}$ steps)
so they were discarded
in the fit. Blue line is the fit result.\label{relaxation}}
\end{figure}

\subsection{The replica method} The Replica method provides us with a standard procedure to compute the complexity and also the in-state entropy \cite{monasson}. 
Its concrete application to hard-sphere systems is described in full-detail in section III of \cite{zamponi}, here we recall it briefly.
It consists in introducing $m$ independent replicas of the system and forcing them to occupy the same metastable state. The entropy of the replicated system becomes then
$$
S(m,\varphi) = \Sigma(\varphi,s) + ms(\varphi)
$$
where $s$ is the free entropy of the state. In the thermodynamic limit, the partition function will be dominated with probability 1 only by the states with the entropy $s_{eq}$ that satisfies the optimum condition
\begin{equation}
\frac{d\Sigma(s,\varphi)}{ds} + m = 0.
\label{eq:seq}
\end{equation}
The in-state entropy $s_{eq}(m,\varphi)$ of those states and their complexity $\Sigma_{eq}(m,\varphi)$ can then be derived using the following relations:
\begin{eqnarray}
s_{eq}(m,\varphi) &=& \frac{\partial S(m,\varphi)}{\partial m},\\
\Sigma_{eq}(m,\varphi) &=& m^2 \frac{\partial [m^{-1}S(m,\varphi)]}{\partial m}, \label{complexityformula}
\end{eqnarray}
And the function $\Sigma(s,\varphi)$ can then be reconstructed from the parametric plots of $s_{eq}(m,\varphi)$ and $\Sigma(m,\varphi)$.

\subsection{Isocomplexity approximation} 
The replica formalism has been applied to the study of infinite dimensional
hard-spheres in the series of papers \cite{denseamorphousHS,denseamorphousHS2,CPUZIII}, with remarkable success. 
However, those results concern only the properties of the glass-former after equilibration, while our numerical results concern the glass former when it is still 
trapped inside a metastable state, \emph{before} equilibration takes place. 
Indeed, one could argue that, for experimental and practical purposes,
getting predictions for this regime is even more important than the study of the equilibrium solution for infinite waiting times. This program however poses a challenge since in principle it requires to solve the dynamics for different preparation protocols. To this day, the only first-principles dynamical theory for glass formers is the Mode Coupling Theory \cite{MCT}, which performs well near the dynamical transition but notoriously fails at higher densities, forcing one to 
use phenomenological models for the description of the high density (or low temperature) regime, as done by Keys et al. in \cite{KGC}. We 
present here a computation which has the advantage of being both fairly simple and \emph{static} in nature.

Since the system is trapped in a single metastable state during the simulation,
it is clear that its physical properties are determined only by the in-state entropy $s(\varphi)$ of that single state. 
We can easily determine $s_{eq}$ at the beginning of the experiment, when the system is at equilibrium and
it corresponds to $s_{eq}(\varphi_0,1)$, but it is nontrivial to determine it when the density is changed and the system falls out of equilibrium, as \eqref{eq:seq} allows us to compute only quantities related to the states that dominate the partition function. Indeed, we can see that for every density $\varphi$ we can choose $s_{eq}$ at our leisure
simply by appropriately tuning the parameter $m$, but in principle we still have no way of knowing what is actually the state the system is trapped into, i.e. we lack a 
criterion to choose a function $m(\varphi)$ consistent with the requirement that the system remains trapped in a single metastable state \cite{lopatin}.

In order to overcome this difficulty, we assume that every state can be followed in density without any crossings between states,
or bifurcations, or spinodal points \cite{montanariricci}; this means that the number (and thus the complexity) of states
that share the same value $s$ of the in-state entropy is a \emph{conserved quantity} during the experiment,
and can then be used as a label for the states. This method is usually referred to as \emph{isocomplexity} \cite{montanariricci,lopatin}.

In summary, to choose $m(\varphi)$ we impose that
\begin{equation}
\Sigma(m(\varphi),\varphi) = \Sigma(1,\varphi_0) = \Sigma_0 = const.
\label{isocomplexity}
\end{equation}
This assumption is false in most cases. For example it has been recently shown that for infinite-dimensional hard spheres a Full 
Replica Symmetry Breaking (fRSB) scenario holds for sufficiently high density \cite{CPUZIII}, invalidating the isocomplexity hypothesis.
The only exact method to tackle the problem would then be the \emph{state following} approach,
which uses the two-replica potential as central tool \cite{FP}. However, this method is far 
more complex and its application goes beyond the scope of this paper, thus we limit ourselves to the isocomplexity assumption,
referring to \cite{corrado} for the complete state-following computation. 
For a systematic comparison of the different approaches in the context of p-spin glasses, see \cite{krzakalazdeborova}.
\begin{figure}
\centering
\includegraphics[width=\columnwidth, angle=0]{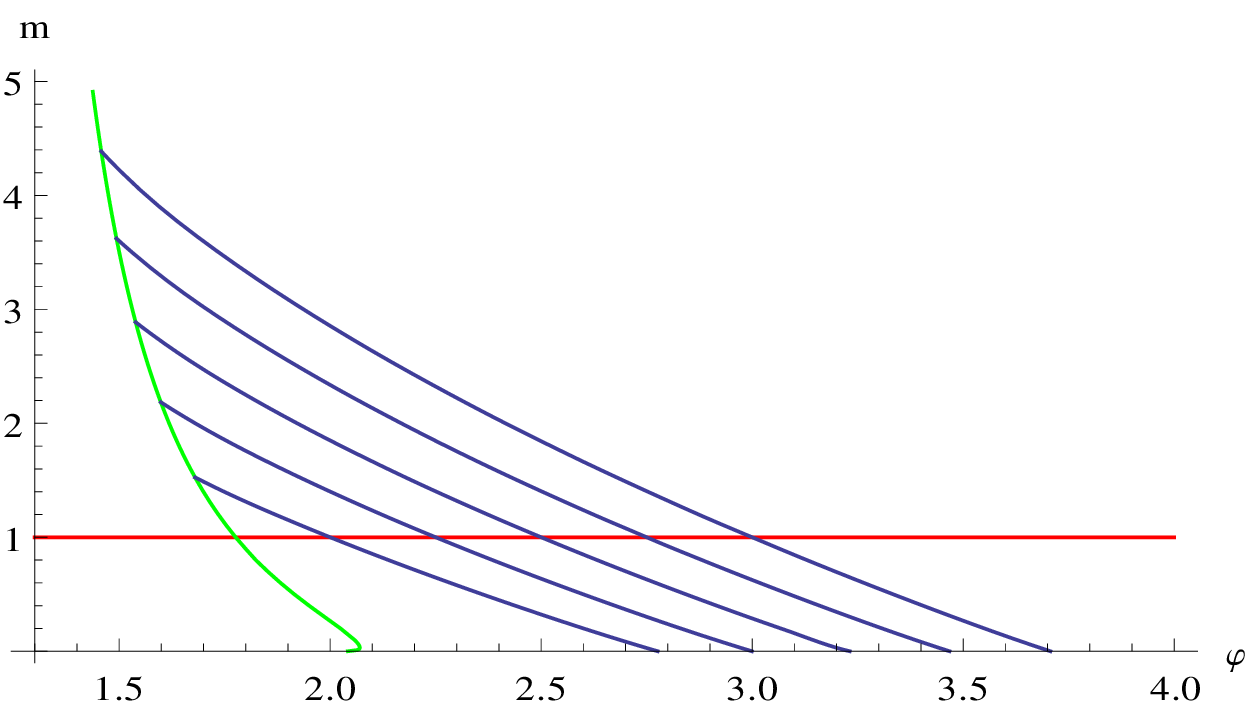}
\caption{Isocomplexity curves (blue lines) in the $(m,\varphi)$ plane for $\varphi_{0}=2,2.25,2.5,2.75,3$. Green line is the clustering line.\label{isocomplexitym}}
\end{figure}


{{We refer to the SI for the details of the computation of the isocomplexity lines displayed in figure \ref{isocomplexitym}.
Once the potential $s_{eq}^{\Sigma_o}(\varphi) = s_{eq}(m_{\Sigma_0}(\varphi),\varphi)$ has been obtained, 
one can compute the desired physical observables using standard thermodynamic relations \cite{mcdonald}. 
Final results for pressure during decompression and compression are shown in Fig. \ref{comparison} and compared to simulation results.
There is a good agreement between analytical and numerical curves, especially for density values not too far from $\varphi_{0}$.}}

\section{Conclusions}
We studied a mean-field model of glass transition, the MK model. We were able both to obtain a stable glass, thanks to the planting technique, and to study numerically and analitically (within replica method and isocomplexity
assumption) the variations of pressure caused by relatively fast changes of density. We showed, both numerically and analitically, that qualitatively this model 
displays the same behaviour of experimental ultrastable 
glasses, reported in \cite{ultrastable, experimental}. 
Our model seems to show a first-order phase transition when evading from metastable equilibrium (see \cite{corrado} and the SI). 
This is in qualitative agreement with the experiments that show that the melting of ultrastable glasses \cite{ultrastable1,ultrastable2} 
has some features in common with first order transitions.

We have also shown that the RFOT approach, together with the replica method, is able to qualitatively describe
the process of glass formation through a slow annealing, with very little computational 
cost and without resorting to \emph{a posteriori} phenomenological considerations.   
Our results can be compared to the DSC experiments where cooling is much slower than heating and as a result the cooled configurations 
(before heating) may be approximated with equilibrium configurations. We can study this situation
in the MK model just because we can plant a thermalised equilibrium configuration at the density we prefer. 
The very interesting problem of understanding the behaviour of DSC experiments when the cooling speed is the same (or faster)
than the heating speed is not studied in this paper: in this situation analytic computations could be done only if we had
under analytic control the dynamics, a goal that has not yet been reached.

\begin{acknowledgments}
The research leading to these results has received funding from the European Research Council under the European Union's Seventh Framework Programme (FP7/2007-2013) / ERC grant agreement n° [247328]. We also thank Francesco Zamponi for useful discussions.
\end{acknowledgments}





\bibliographystyle{pnas}









\renewcommand\thefigure{S\arabic{figure}}

\section{Supplementary Information - Simulation details \label{simulationsdetail}}

We present here all simulation details.

\subsection{Planting method} We generate the initial configuration as follows:
\begin{itemize}
 \item We generate randomly the positions $\underline{x}$ of the $N$ spheres, with an uniform probability distribution over the simulation box.
 \item For each sphere $i$, we generate its shifts $\mathbf{l_{ij}}$ with uniform distribution in the simulation box. 
 Each shift $\mathbf{l_{ij}}$ is accepted if and only if 
 \begin{equation}
    \left|\mathbf{x_{i}-x_{j}+l_{ij}}\right|>1,
  \label{control}
 \end{equation}
 else it is generated again until condition \eqref{control} is satisfied.
\end{itemize}
It can be shown \cite{csp}, that in the infinite volume limit this procedure generates a thermalised initial configuration where the annealed average of entropy is equal to the quenched one,
i.e. a configuration in the liquid phase \cite{csp}.

\begin{figure}[t]
\centering
\includegraphics[width=0.3\textwidth,angle=270]{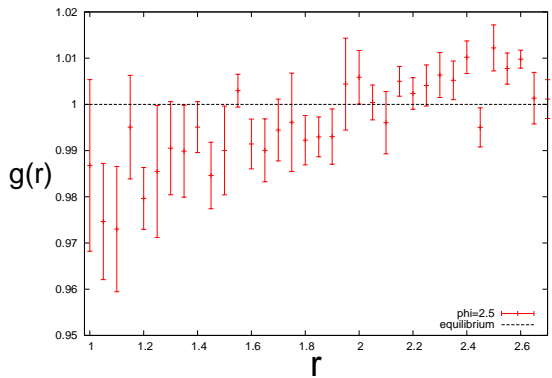}
\caption{Radial distribution function $g(r)$ of the planted configuration at density $\varphi_{0}=2.5$.\label{fig:rdf_zero}}
\end{figure}

We have tested that the procedure works, that the configurations
that we generate are at equilibrium and their properties are independent from time (as long as the density remains constant).
The result for the observable $g(r,0)$ for planting density $\varphi_{0}=2.5$ is shown in figure \ref{fig:rdf_zero}. We can see in the low $r$ region 
a behavior compatible with the
equilibrium one, 
\begin{equation}
 g(\mathbf{x-y})=\exp{(-v(\mathbf{x-y}))}=
 \begin{cases}
  0	&\text{if $|\mathbf{x-y}|<D$}\\
  1	&\text{if $|\mathbf{x-y}|>D$}.
 \end{cases}
 \label{rdfmk}
\end{equation}
within $1$ or $2$ standard deviations. For example for the contact point we have $g(1)=0.987\pm 0.019$. We do not study the large $r$ behaviour of $g(r,0)$,
which shows a decay caused by the finite size of the simulation box and cannot be compared to the infinite-volume analytical result \eqref{rdfmk}.

\subsection{Monte-Carlo evolution algorithm and Verlet lists}
We used a Monte Carlo evolution algorithm. At each step $t$ we propose a displacement $\Delta\mathbf{x_{i}}(t)$ to each sphere $i$.
The proposed displacement is generated uniformly in a $3$-dimensional sphere of radius $\delta$, where $\delta=0.2$ is a fixed parameter.
The proposed displacement is accepted if and only if the condition
\begin{equation}
 \left|\mathbf{x_{i}}+\Delta\mathbf{x_{i}}-\mathbf{x_{j}}+\mathbf{l_{ij}}\right|>1
 \label{control2}
\end{equation}
is satisfied for all other spheres $j\neq i$, else it is rejected. 
This stochastic dynamics satisfies the detailed balance property thus implying relaxation towards equilibrium.
To reduce computational time we use Verlet lists (see for example \cite{understandingmolecular}). 

\subsection{Radial distribution function computation} We denote by $\Omega_{\Delta, \underline{l}}(r, t)$ the number of sphere couples such that
\[
 \left|\mathbf{x_{i}}(t)-\mathbf{x_{j}}(t)+l_{ij}\right|\in [r,r+\Delta]
\]
and we define a fixed time radial distribution function
\[
 g_{\Delta, \underline{l}}(r,t)=\frac{1}{4N\varphi}\frac{\Omega_{\Delta, \underline{l}}(r, t)}{(r+\Delta r)^3-r^3}.
\]
The parameter $\Delta$ is fixed and corresponds to the histogram bin lenght. We choose $\Delta=0.05$.
In order to gain CPU time we perform measurements at equispaced interval in time (typically every 20 Montecarlo sweeps).
In our simulation we further average over the different starting configurations. The number of configurations is $M=6$, a reasonable value for a self-averaging quantity. The statistical error is estimated from sample to sample fluctuations.

\subsection{Decompression protocol} In the following we denote by $k$ the logarithm in base $2$ of the number of Monte-Carlo steps performed for each density value.
We are interested in a decompression protocol that mimics the physical heating of a glass.
We start from a planted initial configuration at a density $\varphi_{0}$ in the glassy region ($\varphi_{0}>\varphi_{d}$) 
in the liquid phase ($\varphi_{0}<\varphi_{k}$), 
equivalent to the supercooled liquid region
in real glass formers. To be safe we choose $\varphi_{0}$ values between $2$ and $3$.
After planting, we decompress the system changing the box size, letting the integer sphere positions unchanged, causing a jump $\varphi_{0}\to\varphi_{1}$
of density, with $\varphi_{1}<\varphi_{0}$. The system evolves for $2^{k}$ Monte-Carlo steps at density $\varphi_{1}$, 
then density jumps again to a lower value $\varphi_{2}<\varphi_{1}$,
the system evolves for $2^{k}$ steps at density $\varphi_{2}$, and so on.

\subsection{Compression protocol} 
To compress the system, we increase the particle radius until the particles touch. When this happens, Montecarlo steps are performed in order to separate the particles; afterwards the radius is increased again, until the final density is reached.\\
The procedure is slow and therefore the final system is nearly thermalised. After the final density is reached, we run a long simulation for final thermalization and we take measurements only in the second half of the run.

 \subsection{Mean Square Displacement \label{msd}}
 
For each density value $\varphi$ we measured the relaxation time by fitting the plateau escape region of the MSD $\Delta(t)$ with the power law
\[ \Delta(t)=a(1+ct^{b}).\]
We discarded the fast relaxation region. We considered for each density value only points with $t>2^{11}$. 
We defined the relaxation time using the relation
$\Delta(\tau_{R})=1.5 \,\Delta(2^{11})$.
The value $1.5$ is somewhat arbitrary: it should be neither too small, to reduce noise effects, neither too large, 
to allow us to obtain relaxation time values not too large compared
to the typical time scales of our simulations. 
Using this procedure we obtained the value of relaxation time for each value of density.
For the highest studied values of density, i.e. $\varphi=1.7,1.68,1.66$, we obtained $\tau_{R}>2^{22}$, meaning that $2^{22}$ steps were 
not sufficient to observe relaxation: these are extrapolated points and we discarded them (red points in Fig. \ref{relaxation}).

\section{Supplementary Information - Analytic computation details}

\subsection{Computation of isocomplexity lines}

The first step is the computation of the replicated entropy $S(m,\varphi)$ as a functional of the replicated density.
Denoting by $\underline{x}=\{\mathbf{x}^{(a)}\}$ the set of the positions of the $m$ replicas, by $\rho(\underline{x})$ the replicated density,
for the mean field MK model we have
\begin{equation}
\begin{split}
 S(m,\varphi)[\rho]= &-\int d\underline{x}\,\rho(\underline{x})\,
 \log{(\rho(\underline{x}))}\\
 &+\frac{1}{2}\int d\underline{x}\,\underline{y}\,\rho(\underline{x})\rho(\underline{y})f(\underline{x}-\underline{y})+N\log{(N)},
 \end{split}
 \label{replicatedentropy}
\end{equation}
where
\[
 f(\underline{x}-\underline{y})=\overline{\exp{\Biggl(\sum_{a=1}^{m}v(\mathbf{x}^{(a)}-\mathbf{y}^{(a)}+\mathbf{l})\Biggr)}}-1
\]
is the replicated Mayer function.
In practice, the replicated density is usually parametrized as
\begin{equation}
 \rho(\underline{x})=\frac{N}{V}\int d\mathbf{X}\,\prod_{a=1}^{m}g_{A}(\mathbf{x}^{(a)}-\mathbf{X}),
 \label{gaussianansatz}
\end{equation}
where $\mathbf{x}^{(a)}$ is the position of replica $a$, and $g_{A}$ is the gaussian function with variance $A$. The parameter $A$ represents the average cage radius and
can be also interpreted as the plateau value of the Mean Square Displacement of particles in the caging regime. For the mean-field MK model of hard spheres, 
combining results presented in Appendix A of \cite{mk} and in section VI of \cite{zamponi}, and putting parametrisation \eqref{gaussianansatz} in \eqref{replicatedentropy}
we obtain
\begin{equation}
\begin{split}
  S(m,\varphi,A)=&\log{(N)}-\log{\Biggl(\frac{\varphi}{\mathcal{V}_{d}(1)}\Biggr)}\\
  &+S_{harm}(m,A)-2^{d-1}\varphi(1-\mathcal{G}(m,A)),
 \label{mkreplica}
 \end{split}
\end{equation}
where $\mathcal{G}(m,A)$ and $S_{harm}(m,A)$ are defined in \cite{zamponi}.
We stress that equation \eqref{mkreplica} is the same of the pure hard-sphere system (without shifts) in infinite dimension.
We must then optimize this with respect to $A$ \cite{zamponi}, getting the equation
\begin{equation}
 \frac{1}{\hat{\varphi}}=\mathcal{F}(m,A(m,\varphi)),
 \label{maximumpointmk}
\end{equation}
where $\hat{\varphi}=2^{d}\varphi/d$ and
\begin{equation}
 \mathcal{F}(m,A)=\frac{A}{1-m}\frac{\partial\mathcal{G}(m,A)}{\partial A}
\label{effe}
\end{equation}

Equation \eqref{maximumpointmk} and the form of the function $\mathcal{F}$ \cite{zamponi} imply a first-order transition at the endpoint of metastable curves, both in the $(m,\varphi)$ plane and in
the pressure-density plane. We discuss this result in the following.\\
We plug the solution $A(m,\varphi)$ of \eqref{maximumpointmk} in \eqref{mkreplica}, obtaining $S(m,\varphi)$.
Using then the replica relations
\begin{eqnarray}
s_{eq}(m,\varphi) &=& \frac{\partial S(m,\varphi)}{\partial m},\\
\Sigma_{eq}(m,\varphi) &=& m^2 \frac{\partial [m^{-1}S(m,\varphi)]}{\partial m},
\end{eqnarray}
on \eqref{mkreplica} we get the following expression for the complexity:
\begin{equation}
\begin{split}
 \Sigma(m,\varphi,A)=& S(m,\varphi,A)\\
 &-\frac{d}{2}(1+m+m\log{(2\pi A)})+2^{d-1}\varphi\mathcal{H}(m,A),
 \label{mkcomplexity}
 \end{split}
\end{equation}
where
\[
 \mathcal{H}(m,A)=-m\frac{\partial \mathcal{G}(m,A)}{\partial m}
\]
The only remaining task is now to solve the equation
$$
\Sigma(m,\varphi) = \Sigma_0 = \Sigma(1,\varphi_0)
$$
with respect to $m$, for various values of $\varphi$. Since in the clustering region $\Sigma$ is a decreasing function of $m$ at fixed $\varphi$, the solution $m_{\Sigma_0}(\varphi)$ of the isocomplexity condition can be found with a simple bisection algorithm. We start at $\varphi_0$, then we change of a small amount $\Delta\varphi$, and we use bisection to find the solution $m_{\Sigma_0}$ of equation
\begin{equation}
\Sigma(m_{\Sigma_0},\varphi_0 + \Delta\varphi) = \Sigma_0, 
\label{isocomplexity}
\end{equation}
Once it has been found, we change the density again and the procedure is repeated until the clustering line is reached and the solution for $A$ disappears.

\subsection{In-state pressure}
In principle the ratio $p(\varphi)$ between physical pressure $P(\varphi)$ of a state of complexity 
$\Sigma_{0}$ and density $\varphi$ can be computed using the relation
\begin{equation}
 p(\varphi)=\rho^{-1}P(\varphi)=-\varphi\frac{d}{d\varphi}s_{eq}(m(\varphi),\varphi),
 \label{pressure}
\end{equation}
where $m(\varphi)$ solves \eqref{isocomplexity} for a given complexity value $\Sigma_{0}$. 
Equation \eqref{pressure} is uncomfortable since it involves also the partial derivative with respect to $m$.
Instead of using directly \eqref{pressure}, we define a modified replicated entropy for each complexity value $\Sigma_{0}$:
\[
 \tilde{S}(m,\varphi)=S(m,\varphi)-\Sigma_{0}=ms_{eq}(m,\varphi)+\Sigma_{eq}(m,\varphi)-\Sigma_{0}.
\]
Isocomplexity equation \eqref{isocomplexity} is then equivalent to the equation
\begin{equation}
 m^{2}\frac{\partial}{\partial m} \Bigl(m^{-1}\tilde{S}(m,\varphi)\Bigr)\Bigg|_{m=m(\varphi)}=0.
\label{isocomplexity1}
\end{equation}
Therefore the pressure of a metastable state can be expressed in terms of total entropy $S$:
\begin{equation}
\begin{split}
 p(\varphi)=&-\frac{\varphi}{m(\varphi)}\frac{\partial}{\partial \varphi}(\tilde{S}(m,\varphi))\Bigg|_{m=m(\varphi)}\\
 =&
 -\frac{\varphi}{m(\varphi)}\frac{\partial}{\partial \varphi}(S(m,\varphi))\Bigg|_{m=m(\varphi)}.
 \end{split}
 \label{pressure_entropy}
\end{equation}
Equation \eqref{pressure_entropy} is all we need to pass from $(m,\varphi)$-plane to pressure-density plane. It is also easy to pass to the contact value of radial distribution function through the relation \cite{mcdonald} $g(1)=(p-1)/(4\varphi)$.

\subsection{Singularity of cage radius and pressure at the clustering line}
\label{singularity}

For each metastable curve $m(\varphi)$ the clustering point $\varphi_{d}$ is defined as the lowest $\varphi$ value for which equation 
\begin{equation}
 \frac{1}{\hat{\varphi}}=\mathcal{F}(m,A(m,\varphi)),
 \label{maximumpointmk}
\end{equation}
admits a finite solution $A(m(\varphi),\varphi)$. The corresponding cage radius $A_{max}$ is the value of $A$ for which $\mathcal{F}(m(\varphi_d),A)$ has a maximum \cite{zamponi}. 
Expanding $\mathcal{F}(m,A)$ in Taylor power series near $\varphi_{d}$ and rearranging terms we obtain from \eqref{maximumpointmk}
\begin{equation}
 A(\varphi)-A_{max}=C\sqrt{\hat{\varphi}-\hat{\varphi}_{d}}+\mathcal{O}(\varphi-\varphi_{d}),
\label{cageradius}
\end{equation}
where $A_{max}=A(m(\varphi_{d}),\varphi_{d})$ and the constant $C$ is given by:
\[
 C=\sqrt{-\frac{1}{2}\frac{\partial^{2} \mathcal{F}(m(\varphi_{d}),A_{max})}{\partial A^{2}}\Biggl( \frac{1}{\hat{\varphi_{d}}^{2}}
 +\frac{\partial\mathcal{F}(m(\varphi_{d}),A_{max})}{\partial m}\frac{\partial m(\varphi_{d})}{\partial\hat{\varphi}}  \Biggr) }.
\]
Equation \eqref{cageradius} implies that 
\begin{equation}
 \frac{dA(\varphi)}{d\varphi}=-\frac{C/2}{(\hat{\varphi}-\hat{\varphi}_{d})^{1/2}}+\text{regular terms},
\end{equation}
i.e. $A'(\varphi)$ has a square-root singularity at $\varphi=\varphi_{d}$.

We show now that this square-root singolarity is transmitted to compressibility.
Expanding the expression for the pressure
\begin{equation}
 p(\varphi)=\frac{1}{m(\varphi)}\biggl(1+2^{d-1}\varphi\Bigl(1-\mathcal{G}(m_{d},A(m(\varphi;\Sigma_{0}),\varphi))\Bigr)\biggr),
 \label{pressurecurves}
\end{equation}
in $\varphi=\varphi_{d}$ we obtain
\begin{equation}
 p(\wh\varphi)=p(\wh\varphi_{d})-B(\wh\varphi)\sqrt{\wh\varphi-\wh\varphi_{d}}+\mathcal{O}(\wh\varphi-\wh\varphi_{d}),
\label{pressuresingularity}
\end{equation}
where we defined the (positive) constant
\[
 B(\wh\varphi)=-\frac{d(1-m(\wh\varphi))}{2m(\wh\varphi)A_{max}}\frac{\wh\varphi}{\wh\varphi_{d}}C.
\]
Deriving equation \eqref{pressuresingularity} we obtain
\begin{equation}
 \frac{dp(\wh\varphi)}{d\wh\varphi}=-\frac{B(\wh\varphi_{d})/2}{(\hat{\varphi}-\hat{\varphi}_{d})^{1/2}}+\text{regular terms},
 \label{pressuresingularity1}
\end{equation}
i.e. the derivative $p'(\wh\varphi)$ of the pressure has a singularity in $\wh\varphi=\wh\varphi_{d}$ with the same critical exponent of $A'(\wh\varphi)$. 
This fact implies an overshoot in the pressure as the system escapes from the metastable state. 
Indeed, the same overshoot can be seen also in the state following method \cite{corrado}, not only in the pressure vs. density 
plane, but also in the shear stress vs. shear strain plane.









\end{article}

\end{document}